\documentclass[12pt]{iopart}
\usepackage{graphicx,rotating,subfigure,delarray,color}
\usepackage{iopams}
\expandafter\let\csname equation*\endcsname\relax
\expandafter\let\csname endequation*\endcsname\relax
\usepackage{amsmath}
\usepackage{hyperref}
\usepackage{xcolor}
\usepackage{soul}
\usepackage{dsfont}
\usepackage[T1]{fontenc}
\usepackage{physics}
\usepackage{multirow}
\usepackage{bbm}
\usepackage{tabularx}
\usepackage{setspace}
\usepackage[export]{adjustbox}
\renewcommand{\vec}[1]{\boldsymbol #1}
\newcommand{\im}{\text{i}}
\def\l{\left}
\def\r{\right}
\def\12{\frac{1}{2}}

\begin{document}
\title[Symmetry-Resolved Entanglement: General considerations, ...]{Symmetry-Resolved Entanglement: General considerations, calculation from correlation functions, and bounds for symmetry-protected topological phases}
\author{Kyle Monkman and Jesko Sirker}
\address{Department of Physics and Astronomy and Manitoba
Quantum Institute, University of Manitoba, Winnipeg, Canada R3T 2N2}
\date{\today}
\begin{abstract}
We discuss some general properties of the symmetry-resolved von-Neumann entanglement entropy in systems with particle number conservation and describe how to obtain the entanglement components from correlation functions for Gaussian systems. We introduce majorization as an important tool to derive entanglement bounds. As an application, we derive lower bounds both for the number and the configurational entropy for chiral and $C_n$-symmetric topological phases. In some cases, our considerations also lead to an improvement of the previously known lower bounds for the entanglement entropy in such systems. 
\end{abstract}
\vspace{2pc}

\maketitle

\section{Introduction}
In the presence of particle number conservation, the reduced density matrix $\rho_A$ of a system $S=A\cup B$ has block structure, i.e., $[\rho_A,N_A]=0$ where $N_A$ is the particle number operator for subsystem $A$.  As a consequence, the von-Neumann entanglement entropy can be rewritten in a symmetry-resolved manner
\begin{equation}
\label{Eq1}
S[\rho_A] = -\tr\rho_A\ln\rho_A = -\sum_n \tr \rho_n\ln\rho_n \, .
\end{equation}
Here $\rho_n$ is the block of $\rho_A$ with particle number $n$. Studying the symmetry-resolved entanglement has been a subject of recent interest. One of the main results of these studies is that in one-dimensional critical systems, there is an equipartition of the entanglement between the different symmetry sectors 
\cite{Xavier,CFTCalabrese}. One can, furthermore, also separate the entanglement entropy for a system with particle number conservation into two distinct components. To do so, one can write $\rho_n=p_n\tilde\rho_n$ with $p_n=\tr\rho_n$ being the probability to find $n$ particles in subsystem $A$ and $\tr\tilde\rho_n=1$. Plugging this expression into Eq.~\eqref{Eq1} leads to 
\begin{equation}
\label{Eq2}
S[\rho_A] = \underbrace{-\sum_n p_n\ln p_n}_{S_N} \underbrace{+\sum_n p_n S[\tilde\rho_n]}_{S_c} \, .
\end{equation}
The first part is the Shannon entropy of the particle number distribution which is called the number entropy $S_N$ while the second part is the configurational entropy $S_c$ \cite{Wiseman,Greiner,kiefer2020bounds,MonkmanSirkerEdge}. The number entropy is the part of the entanglement which can often be accessed experimentally without requiring a full quantum tomography of the state \cite{Greiner}. The configurational entropy, on the other hand, is a special case of the operational entanglement entropy---first introduced by Wiseman and Vacaro \cite{Wiseman} as the entanglement extractable from a quantum many-body system of indistinguishable particles and transferable to a quantum register---for the case of a bipartition of a pure quantum state. 
The separation of the entanglement entropy in the presence of particle number conservation into these two components has been used to study study many-body physics \cite{KieferSirker1, KieferSirker2, kiefer2020bounds, Greiner, Parez, Bonsignori, Sela}, quantum field theories \cite{field1, field2, field3, field4, field5}, and topological systems \cite{MonkmanSirkerEdge, topology1, topology2}.

The fact that the von-Neumann entropy is extensive is a fundamental result of Statistical Mechanics. However, the same is not true for the symmetry-resolved components. As we will show here, it is nevertheless possible to derive inequalities for the number and configurational entropies of general subsystems. For Gaussian systems, these inequalities imply that minimal bounds on the entanglement can be found by considering only a subset of the single-particle entanglement eigenvalues. To calculate such bounds, we express the number and the configurational entropy for fermionic Gaussian systems in terms of the eigenvalues of the correlation matrix for the subsystem $A$. This is a generalization of the methods developed by Peschel \cite{peschel} to the symmetry-resolved case.

Optimization via majorization is often used for entangled systems since entropy is a concave function \cite{nielsen2002quantum,nielsen2001majorization}. Previously known was that majorization of the single-particle entanglement spectrum minimizes the von-Neumann entanglement entropy. Here we use the Shepp-Olkin Majorization Theorem \cite{SheppOlkin, Olkin, HillionJohnson} to show that the number entropy is also a concave function of the single-particle entanglement eigenvalues. For the configurational entropy we hypothesize, based on results for small subsets of eigenvalues, that it is concave in the single-particle eigenvalues as well. Thus majorizing the single-particle entanglement spectrum not only minimizes the von-Neumann entanglement entropy but also the number and configurational entropies.

As an application, we use these methods to find lower entanglement bounds both for chiral insulators \cite{MonkmanSirkerSpectrum} as well as for $C_n$ symmetric topological crystalline insulators \cite{Bernevig2011,Bernevig2013,Bernevig2014}. The latter extends previously known results for lower bounds on the symmetry-resolved entanglement in $C_2$-symmetric topological insulators \cite{C2MonkmanSirker}. For certain cuts of $C_n$-symmetric systems we find, using majorization, additional restrictions on the entanglement spectrum beyond what is currently known. In addition to novel bounds on the number and configurational entropy, these restrictions imply also a new, stronger bound on the von-Neumann entanglement entropy. 

Our paper is organized as follows: In Sec.~\ref{section:SRE}, we prove some general inequalities for the symmetry resolved entanglement valid for any system. In Sec.~\ref{section:FGS}, we then specifically consider fermionic Gaussian systems, express the symmetry-resolved entanglement components in terms of the eigenvalues of the correlation matrix, and discuss majorization techniques. This allows us to establish lower bounds for the symmetry-resolved entanglement of chiral and $C_n$-symmetric topological insulators which are discussed in Sec.~\ref{section:CN}. In Sec.~\ref{section:CNexamples}, we analyze these bounds for specific examples of $C_4$-symmetric insulators. The final section summarizes the obtained results and provides an outlook on some of the remaining open questions. 

\section{Symmetry-Resolved Entanglement}
\label{section:SRE}
The von-Neumann entanglement entropy, number entropy, and configurational entropy are all functions of some density matrix $\rho$, see Eq.~\eqref{Eq2}. In this paper, we will be interested in the case where $\rho$ is a reduced density matrix obtained after splitting a system into two subsystems and tracing out one of them. However, for the following general considerations this does not matter and $\rho$ can be any density matrix.

Let us assume that $\rho=\rho^X\otimes\rho^Y$ with $[\rho^X,N^X]=[\rho^Y,N^Y]=[\rho,N]=0$ where $N^X$ and $N^Y$ are the particle number operators for $X$ and $Y$, respectively, and $N=N^X+N^Y$. Furthermore, $\rho^X$ and $\rho^Y$ are proper density matrices, in particular, $\tr\rho^X=\tr\rho^Y=1$. For the von-Neumann entropy it follows that 
\begin{eqnarray}
    \label{vN}
    S &=&  -\tr\rho\ln\rho =-\tr\l\{ \l(\rho^X\otimes\rho^Y\r)\ln\l(\rho^X\otimes\rho^Y\r)\r\} \nonumber \\
    &=& -\tr\l\{ \l(\rho^X\otimes\rho^Y\r)\ln\rho^X\r\}-\tr\l\{ \l(\rho^X\otimes\rho^Y\r)\ln\rho^Y\r\} \\
    &=& S[\rho^X] + S[\rho^Y] \nonumber
\end{eqnarray}
which is its well-known extensivity property. However, the symmetry-resolved components \eqref{Eq2} are {\it not extensive}. In the following we will show that 
\begin{equation}
\label{SCSNresult}
    S_c[\rho] \geq S_c[\rho^X]+S_c[\rho^Y], \qquad
    S_N[\rho] \geq \mbox{max}\lbrace S_N[\rho^X],S_N[\rho^Y] \rbrace.
\end{equation}
To show these relations, we start by noting that---according to our assumptions---$\rho$, $\rho^X$, and $\rho^Y$ all have block structure with respect to their respective particle numbers. We have, in particular, 
\begin{equation}
\label{rhon}
\tilde\rho_{n} = \sum_{r=0}^{n} \frac{p_{r}^X p_{n-r}^Y}{p_n} \  \tilde\rho_{r}^X \otimes \tilde\rho_{n-r}^Y, \qquad
p_n = \sum_{r=0}^{n} p_r^X p_{n-r}^Y
\end{equation}
where $\tilde\rho_n$ denotes normalized density matrices with $\tr\tilde\rho_n=1$ and $p_n$ is the probability to have $n$ particles. We can diagonalize the blocks $\tilde\rho^X$ and $\tilde\rho^Y$ at the same time, implying that the eigenvalues of $\tilde\rho_n$ are given by
\begin{equation}
    \label{rho_eval}
    \lambda_r(i,j)= \frac{p_{r}^X p_{n-r}^Y}{p_n}\lambda_r^X(i)\lambda_{n-r}^Y(j) 
\end{equation}
for $r=0,\dots,n$. For the configurational entropy, this implies that
\begin{eqnarray}
    \label{Sc_proof}
    S_c[\rho] &=&-\sum_n p_n\tr\tilde\rho_n\ln\tilde\rho_n 
    =-\sum_{n,r,i,j}p_n\lambda_r(i,j)\ln\lambda_r(i,j)\nonumber \\
    &\geq& -\sum_{n,r,i,j} p_{r}^X p_{n-r}^Y \lambda_r^X(i)\lambda_{n-r}^Y(j) \ln\l(\lambda_r^X(i)\lambda_{n-r}^Y(j) \r) \\
    &=& S_c[\rho^X] + S_c[\rho^Y] \nonumber
    \end{eqnarray}
    where we have used $\frac{p_{r}^X p_{n-r}^Y}{p_n} \leq 1$. This proves the configurational entropy bound \eqref{SCSNresult}. 

Now we focus on showing the inequality \eqref{SCSNresult} for the number entropy $S_N$. Defining a concave function $f(p)=-p \ln(p)$, our goal is to demonstrate that
\begin{equation}
\label{SNbound}
S_N[\rho]=\sum_{n=0}^M f(p_n) \geq S_N[\rho^X]=\sum_{n=0}^M f(p_n^X).
\end{equation}
If we can show the above inequality, then it will also be true for $S_N[\rho^Y]$ and thus for the maximum of the number entropies of the two subsystems. Since $f(p)$ is concave, we can use Karamata's inequality \cite{Karamata1,Karamata2,Karamata3}. Let $\sigma(i)$ and $\alpha(i)$ be permutations of $(0, \ 1, \  \dots \ M)$ such that
$x_0=p_{\sigma(0)}^X\geq x_1=p_{\sigma(1)}^X\geq\dots\geq x_M=p_{\sigma(M)}^X$ and $y_0=p_{\alpha(0)}\geq y_1=p_{\alpha(1)}\geq\dots\geq y_M=p_{\alpha(M)}$. If we can show that $\vec{x}=(x_0, \ x_1, \  \dots \ x_M)$ majorizes $\vec{y}=(y_0, \ y_1, \  \dots \ y_M)$, then \eqref{SNbound} is true by Karamata's inequality. Since they are probability distributions, we know that $\sum_{i=0}^M x_i = \sum_{i=0}^M y_i = 1$. We now simply need to show that $\sum_{i=0}^s y_i \leq \sum_{i=0}^s x_i$ for all $s =  0, 1, \dots M-1$. So
\begin{eqnarray}
\sum_{i=0}^s y_i &=& \sum_{i=0}^s p_{\alpha(i)}
=\sum_{i=0}^s \sum_{j=0}^{\alpha(i)} p_j^Y p_{\alpha(i)-j}^X 
=\sum_{j=0}^M p_j^Y (\sum_{i=0}^s p_{\alpha(i)-j}^X)  \nonumber \\
&\leq& 
\sum_{j=0}^M p_j^Y (\sum_{i=0}^s x_i)  = \sum_{i=0}^s x_i. 
\end{eqnarray}
Therefore $\vec{x}$ majorizes $\vec{y}$. Thus the inequality for the number entropy \eqref{SCSNresult} has been proven.

\section{Fermionic Gaussian Systems}
\label{section:FGS}
It is well known that for Gaussian systems all properties of a subsystem can be calculated from the correlation matrix $C$ of the subsystem with matrix elements $C_{nm}=\langle c_n^\dag c_m\rangle$ \cite{peschel}. Here we will focus on fermionic Gaussian systems with Hamiltonian $H=-\sum_{n,m} t_{nm} c_n^\dag c_m$ where $t_{nm}$ are hopping amplitudes and $c_n$ fermionic annihilation operators. We note that it should be possible to generalize the results below to bosonic Gaussian systems as well. 

\subsection{Reduced density matrix and correlation functions}
Following Ref.~\cite{peschel}, we know that for fermionic Gaussian systems a reduced density matrix of a subsystem with $M$ sites can be written as
\begin{equation}
    \label{RDM_free}
    \rho=\frac{\text{e}^{-\mathcal{H}}}{Z}=\frac{1}{Z}\exp\l(-\sum_{k=1}^M \varepsilon_k a_k^\dag a_k\r), \, Z=\prod_{k=1}^M \l(1+\text{e}^{-\varepsilon_k}\r) . 
\end{equation}
The eigenvalues $\varepsilon_k$ of the so-called bilinear entanglement Hamiltonian $\mathcal{H}$ can be related to the eigenvalues $C_k$ of the correlation matrix $C$ by
\begin{equation}
    \label{eps_to_C}
    \exp(-\varepsilon_k)=\frac{C_k}{1-C_k} \, .
\end{equation}
From Eq.~\eqref{RDM_free} it follows that we can write $\rho$ as a diagonal $2^M\times 2^M$ matrix 
\begin{equation}
    \label{RDM_blocks}
    \rho=\frac{1}{Z}\bigotimes_{k=1}^M \l(\begin{array}{cc} 1 & 0 \\ 0 & \text{e}^{-\varepsilon_k} \end{array}\r)=\bigotimes_{k=1}^M \l(\begin{array}{cc} 1-C_k & 0 \\ 0 & C_k \end{array}\r) \, .
\end{equation}
Thus, the $2^M$ eigenvalues of $\rho$ are given by
\begin{equation}
    \label{EV_free}
 \lambda_{\{n_k\}} =\frac{1}{Z}\prod_{k=1}^M \l(\text{e}^{-\varepsilon_k}\r)^{n_k} =  \prod_{k=1}^M C_k^{n_k}(1-C_k)^{1-n_k}
\end{equation}
where $\{n_k\}$ is a list of length $M$ with $n_k=0,1$. The $\binom{M}{n}$ eigenvalues in the n-particle block of the density matrix can then be written as
\begin{equation}
    \label{EV_block}
    \lambda^{(n)}_j = C_{s_j(1)}\dots C_{s_j(n)}(1-C_{s_j(n+1)})\dots (1-C_{s_j(M)}) 
\end{equation}
where $s_j$ are permutations of the numbers $\{1,\dots,M\}$ with $s_j(1)<\dots<s_j(n)$ and $s_j(n+1)<\dots<s_j(M)$. The probability $p_n$ to find $n$ particles in the subsystem is then given by 
\begin{equation}
    \label{RDM_probs}
    p_n = \sum_j \lambda^{(n)}_j 
    = \sum_j C_{s_j(1)}\dots C_{s_j(n)}(1-C_{s_j(n+1)})\dots (1-C_{s_j(M)}) 
\end{equation}
where the same conditions on the permutations $s_j$ apply as before. In particular, $p_0=Z^{-1}=\prod_{k=1}^M(1-C_k)$ and $p_M=\prod_{k=1}^M C_k$. We note that Eq.~\eqref{RDM_probs} is the probability mass function of the Poisson binomial distribution, i.e., it can be interpreted as the probability to have $n$ successful tries (particle present) out of a total of $M$ tries. Since the sum in Eq.~\eqref{RDM_probs} contains $\binom{M}{n}$ terms, this sum is impractical to calculate for large $M$. For the sake of compeleteness, we note that the probability mass function $p_n$ of the Poisson binomial distribution can be more efficiently calculated by a discrete Fourier transform
\begin{equation}
    \label{Fourier}
    p_n=\frac{1}{M+1}\sum_{l=0}^M T_l^{-n}\prod_{m=1}^M (1+(T_l-1)p_m)
\end{equation}
with $T_l=\exp(-2\pi \im l/(M+1))$. This approach of calculating the probabilities $p_n$ was used for the number entropy $S_N$ in Ref.~\cite{KieferSirker1}.

Using Eq.~\eqref{EV_free}, we can write the von-Neumann entanglement entropy as \cite{peschel}
\begin{equation}
    \label{SvN}
    S=\sum_k\l\{\ln\l(\text{e}^{-\varepsilon_k}+1\r)+\frac{\varepsilon_k}{\text{e}^{\varepsilon_k}+1}\r\} 
    =-\sum_k\l\{ C_k\ln C_k +(1-C_k)\ln(1-C_k) \r\} \, .
\end{equation}
According to Eq.~\eqref{RDM_probs}, we can express the probabilities as $p_n=p_n(C_1,\dots,C_M)$. Thus, the number entropy $S_N=-\sum_n p_n\ln p_n$ is also a function of the correlation matrix eigenvalues $\{ C_k \}$. The configurational entropy can then be obtained as $S_c=S-S_N$ and is a function of the $\{ C_k \}$ as well.

In terms of deriving bounds for the symmetry-resolved entanglement, we can draw the following conclusions: (i) From Eq.~\eqref{RDM_blocks} we see that we can treat each eigenvalue of the correlation matrix as an 'independent' contribution. I.e., the inequalities \eqref{SCSNresult} are applicable to any subset of eigenvalues. (ii) The symmetry-resolved entanglement components can be expressed explicitly by the correlation matrix eigenvalues $\{C_k\}$. The entanglement entropy \eqref{SvN} is a concave function 
 of $\{C_k\}$. If we can show that $S_N$ and $S_c$ are concave functions of the $\{C_k\}$ as well, then we can apply majorization techniques to derive bounds for all three entanglement measures. We will discuss this approach in the following subsections.

\subsection{Majorization of $S$ and $S_N$}
The entanglement entropy \eqref{SvN} is a sum of concave functions of the $\{C_k\}$. We now define two vectors $\vec{C}=(C_{1},C_{2},\dots, C_{M})$ and $\vec{C}'=(C_{1}',C_{2}',\dots, C_{M}')$, ordered such that $C_1\geq C_2\geq\dots\geq C_M$ and $C_1'\geq C_2'\geq\dots\geq C_M'$. We then say that $\vec{C}'$ majorizes $\vec{C}$ written as $\vec{C}' \succ \vec{C}$ if
\begin{equation}
    \label{part_number}
    \sum_{k=1}^M C_k = \sum_{k=1}^M C_k' =\langle N\rangle \, ,
\end{equation}
where $\langle N\rangle$ is the average particle number in the subsystem and
\begin{equation}
    \label{major}
    \sum_{k=1}^n C_k'\geq \sum_{k=1}^n C_k\quad\mbox{for}\quad n=1,\dots,M-1 \, .
\end{equation}
Under the conditions \eqref{part_number} and \eqref{major}, Karamata's inequality is applicable to the concave entanglement entropy function \eqref{SvN} and we have 
\begin{equation}
    S(C_1',C_2',\dots,C_M') \leq S(C_1,C_2,\dots,C_M)\, .
\end{equation}
Without any additional restrictions on the eigenvalues of the correlation matrix, this implies the intuitive result that the maximum entropy is obtained when $C_k=\langle N \rangle / M$ for all $k$. The minimum entropy $S=0$ can be obtained by making each $C_k$ equal to either $0$ or $1$.

Applying majorization to the number entropy is more complicated. While $S_N$ is a sum of concave functions in the probabilities $\{p_n\}$ it is not immediately obvious if it is concave in the $\{C_k\}$ as well. Here we will make use of some more advanced concepts and results in majorization and probability theory. We first note that the number of particles $N$ found in the subsystem in a measurement is a sum of Bernoulli random variables
\begin{equation}
    N=X_1 +X_2+\dots+X_M
\end{equation}
which occur with probabilities $P(X_i=1)=C_i$ and $P(X_i=0)=1-C_i$. The $p_n(\{C_k\})$ are then, as already eluded to earlier, the probability mass function of the Poisson binomial distribution with parameters $\{C_k\}$. The number entropy $S_N$ is the Shannon entropy of this mass function. It has been conjectured by Shepp and Olkin \cite{SheppOlkin,Olkin} and proven by Hillion and Johnson \cite{HillionJohnson} that the Shannon entropy in this case is not only a concave function of the $p_n$ but of the parameters $\{C_k\}$ as well. This result is known as the Shepp-Olkin theorem. We therefore find again that if $\vec{C}' \succ \vec{C}$, we have 
\begin{equation}
    S_N(C_1',C_2',\dots,C_M') \leq S_N(C_1,C_2,\dots,C_M).
\end{equation}
Without further restrictions on the $\{C_k\}$, this implies again that the maximum number entropy is obtained if $C_k=\langle N \rangle / M$ for all $k$ and the minimum number entropy $S_N=0$ if each $C_k$ is equal to either $0$ or $1$. 

\subsection{A conjecture on the concavity of $S_c$}
We note that the configurational entropy $S_c=S+(-S_N)$ is a sum of a concave and a convex function. Therefore, one cannot directly apply majorization to $S_c$ based on the already obtained results for $S$ and $S_N$. However, this does not rule our that $S_c$ is also a concave function of the parameters $\{C_k\}$. Similar to the original Shepp-Olkin paper \cite{SheppOlkin} we will here just investigate the cases $M=2$ and $M=3$. Based on these results, we conjecture that $S_c(\{C_k\})$ is indeed Schur concave and majorization thus applies. 

\subsubsection{The case $M=2$:} 
\begin{figure}
    \includegraphics[width=0.55\columnwidth]{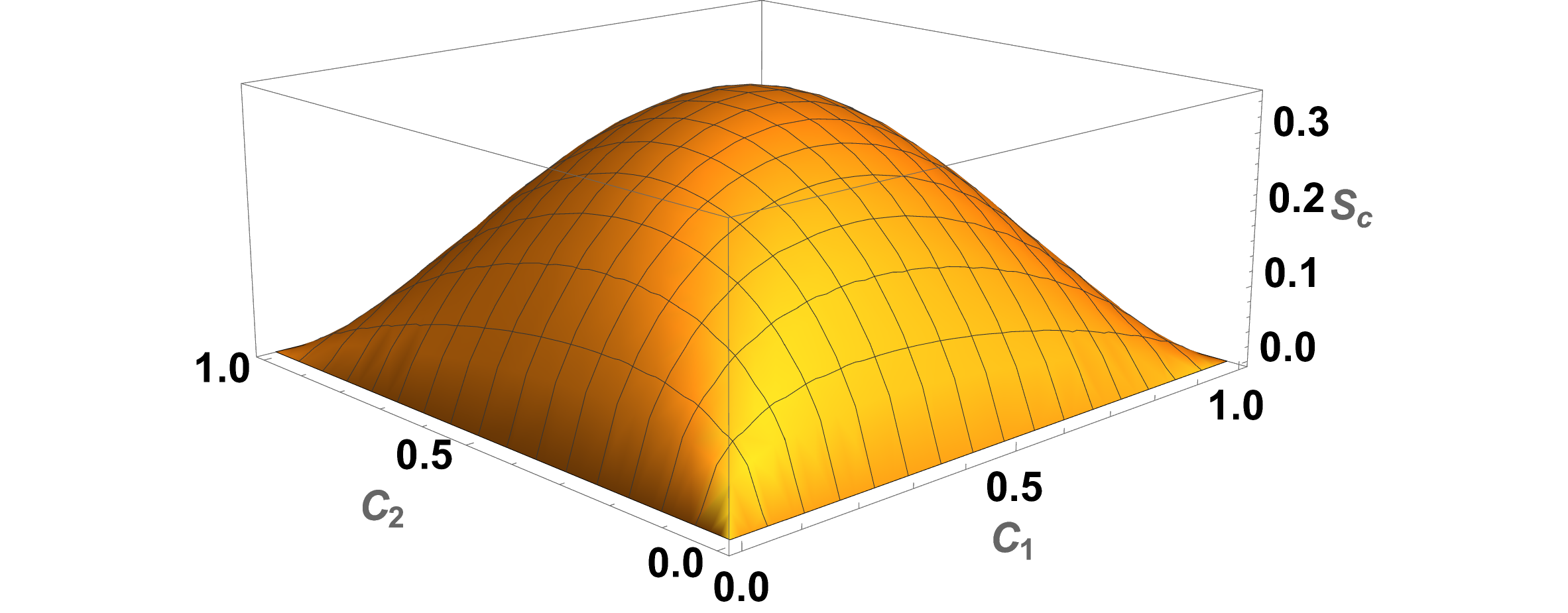}
    \hfill
    \includegraphics[width=0.55\columnwidth]{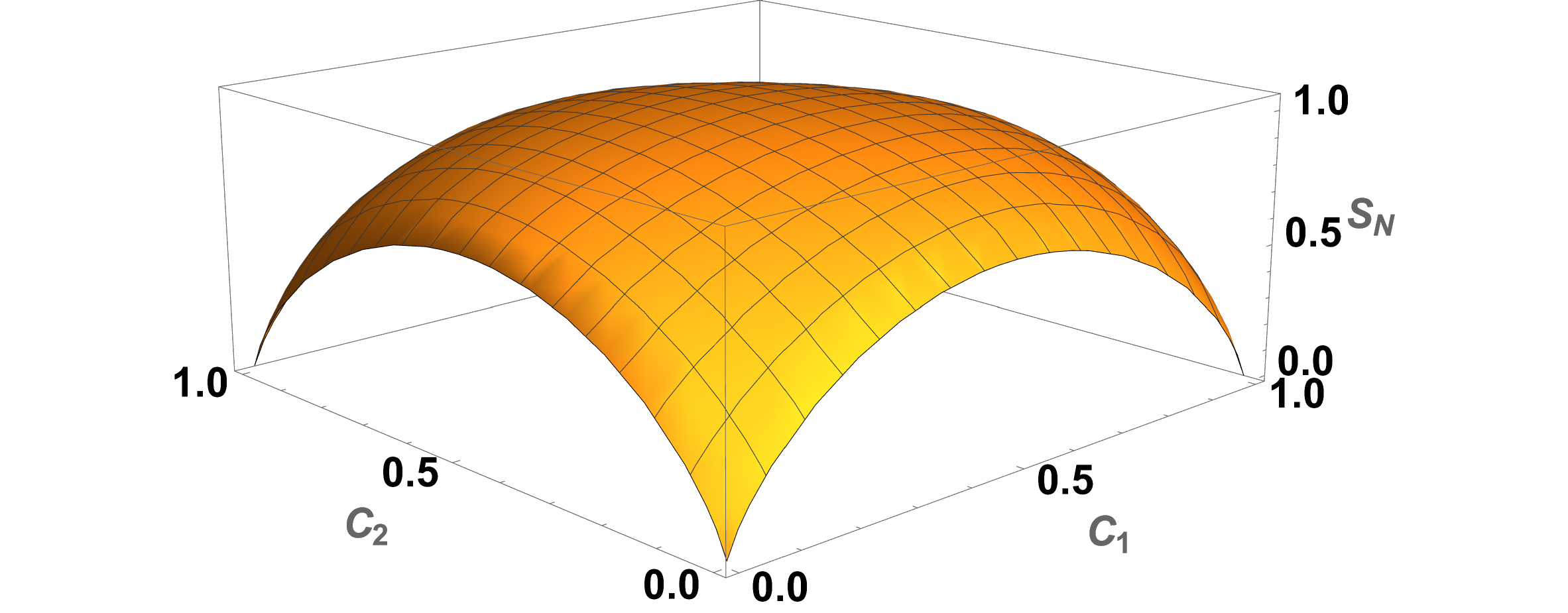}
    \caption{Left panel: The configurational entropy as a function of $C_1$ and $C_2$ for the $M=2$ case is concave. Right panel: The number entropy is also concave as shown earlier also for general $M$.}
    \label{Sc_fig}
\end{figure}

From Eq.~\eqref{EV_block}, we find that in this case the eigenvalues of the reduced density matrix are given by
\begin{eqnarray}
    \label{M2_case}
    \lambda^{(0)}&=&p_0=(1-C_1)(1-C_2), \quad\lambda^{(2)}=p_2=C_1C_2 \\
    \lambda^{(1)}_1 &=& C_1(1-C_2),\quad \lambda^{(1)}_2=(1-C_1)C_2,\quad
    p_1 = C_1(1-C_2)+(1-C_1)C_2 \nonumber
\end{eqnarray}
The configurational entropy is thus given by 
\begin{equation}
    \label{Sc_M2}
    S_c=-\lambda^{(1)}_1\ln(\lambda^{(1)}_1)-\lambda^{(1)}_2\ln(\lambda^{(1)}_2)+p_1\ln p_1 \, .
\end{equation}
 As can be seen in Fig.~\ref{Sc_fig}, the configurational entropy is indeed concave in $\{C_1,C_2\}$.
This can also be checked by using the Schur-Ostrowski criterion
\begin{equation}
    \label{Schur-Ostrowski}
    (C_i-C_j)\l(\frac{\partial{S_c}}{\partial C_i}-\frac{\partial{S_c}}{\partial C_j}\r)\leq 0
\end{equation}
for $1\leq i\neq j\leq M$.
The maximal values for all three entropies for an average filling $\langle N\rangle = 1$ are thus obtained for $(C_1,C_2)=(1/2,1/2)$ and are given by
\begin{equation}
    \label{Smax_M2}
    S=2\ln 2,\;\; S_N=\frac{3}{2}\ln 2,\;\; S_c=\frac{1}{2}\ln 2 \, .
\end{equation}

\subsubsection{The case $M=3$:}  
\begin{figure}
    \includegraphics[width=0.43\columnwidth]{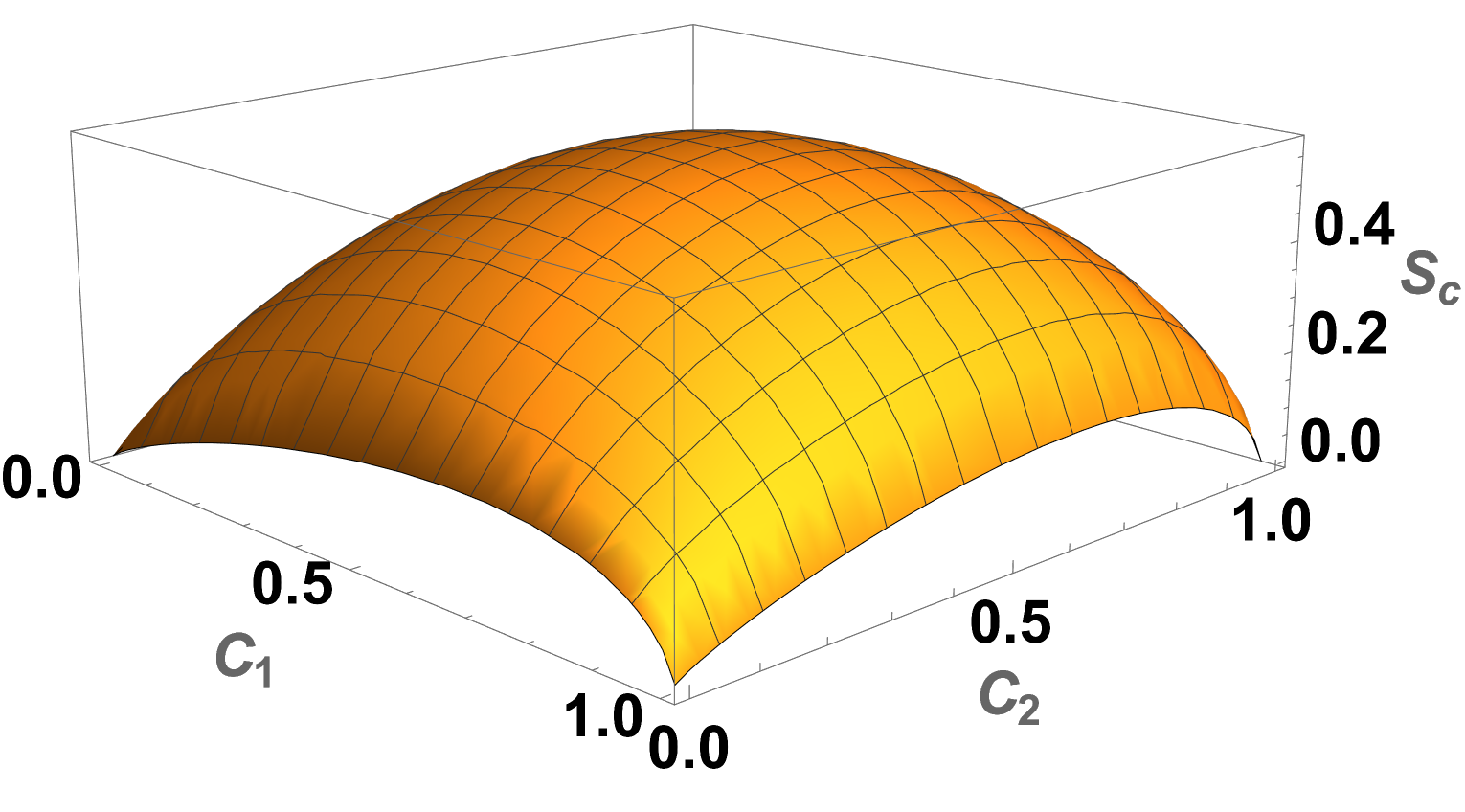}
    \hfill
    \includegraphics[width=0.43\columnwidth]{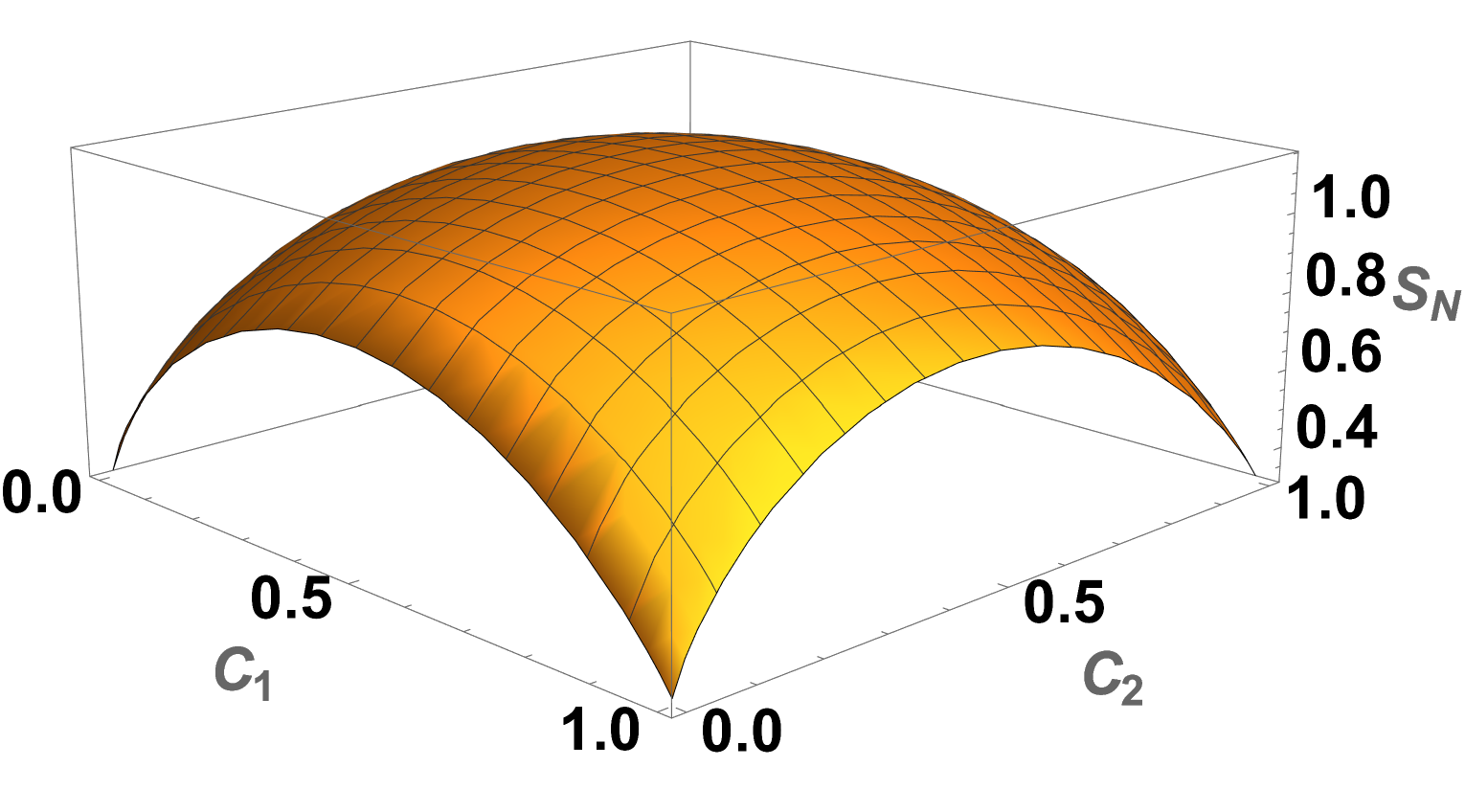}
    \caption{Same as Fig.~\ref{Sc_fig} but for $M=3$ with $C_3=0.9$ fixed.}
    \label{Sc_fig2}
\end{figure}

We introduce the notation $\bar C_k = (1-C_k)$. Then the eigenvalues of the reduced density matrix for $M=3$ are given by
\begin{eqnarray}
    \label{M3_case}
    \lambda^{(0)}&=&p_0=\bar C_1 \bar C_2 \bar C_3,\qquad \lambda^{(3)}=p_3=C_1C_2C_3 \nonumber \\
    \lambda^{(1)}_1 &=& C_1\bar C_2 \bar C_3,\qquad
    \lambda^{(1)}_2=\bar C_1 C_2 \bar C_3,\qquad\lambda^{(1)}_3=\bar C_1 \bar C_2 C_3  \\
    \lambda^{(2)}_1&=&C_1C_2\bar C_3,\qquad \lambda^{(2)}_2=C_1 \bar C_2 C_3,\qquad \lambda^{(2)}_3=\bar C_1 C_2C_3 \nonumber
\end{eqnarray}
with $p_1=\sum_j \lambda_j^{(1)}$ and $p_2=\sum_j \lambda_j^{(2)}$.
We can now calculate $S_N(C_1,C_2,C_3)$ and $S_c(C_1,C_2,C_3)$. Plots of both quantities for a fixed value of $C_3$ are shown in Fig.~\ref{Sc_fig2}.
We find again that $S_c(C_1,C_2,C_3)$ is concave, a conclusion which we also checked using the Schur-Ostrowski criterion \eqref{Schur-Ostrowski}. In the case $M=3$ we then have the following non-trivial maximal values for the entropies: For filling $\langle N\rangle =\frac{3}{2}$, for example, we have
\begin{equation}
    \label{Smax_M3}
    S=3\ln 2,\quad S_N=3\ln 2-\frac{3}{4}\ln 3,\quad S_c=\frac{3}{4}\ln 3 \, ,
\end{equation}
while for fillings $\langle N\rangle =1$ and $\langle N\rangle =2$ we have
\begin{equation}
    \label{Smax_M3_2}
    S=3\ln 3-2\ln 2,\quad S_N=\frac{7}{3}\ln 3-2\ln 2,\quad S_c=\frac{2}{3}\ln 3 \, .
\end{equation}

For the half-filled case, $\langle N\rangle =\frac{M}{2}$, the entropies are always maximized by setting $C_k=\frac{1}{2}$ for $k=1,\dots,M$. This means, according to Eq.~\eqref{EV_free}, that all eigenvalues of the reduced density matrix are equal and given by $\lambda=1/2^M$. Since each $n$-particle block has $\binom{M}{n}$ eigenvalues, the probabilities are given by $p_n=\binom{M}{n}/2^M$. The maximal entropies are therefore given by
\begin{equation}
    \label{Smax_general}
    S=M\ln 2,\quad S_c=\frac{1}{2^M}\sum_{n=0}^M B^M_n \ln B^M_n,\quad S_N=S-S_c \, .
\end{equation}
where we have defined $B^M_n=\binom{M}{n}$.

To briefly summarize: One can use the structure of the reduced density matrix in the Gaussian case, see Eq.~\eqref{RDM_blocks}, together with the inequalities \eqref{SCSNresult} to derive lower bounds for the symmetry-resolved entanglement components based on a subset of the eigenvalues of the correlation matrix. Furthermore, majorization is applicable because all entanglement components are Schur concave functions. 

So far, we have discussed the case where the only restriction on the correlation matrix eigenvalues $C_k$ is the average particle number in the subsystem, $\langle N\rangle = \sum_k C_k$. The maximal entropies for all of $S$, $S_c$, and $S_N$ are then obtained if $C_k=\langle N\rangle/M$ for all $k$. If $\langle N\rangle$ is an integer, then the minimum entropy is $S=0$, obtained by setting the appropriate number of $C_k=1$ and the rest to zero. Interestingly, if $\langle N\rangle$ is not an integer then there will be a non-trivial lower entanglement bound. I.e., such non-integer, average particle numbers are only possible in entangled states. This is an obvious example for a case where there is a non-trivial lower entanglement bound but it does not tell us in which types of systems such bounds exist. 

It is thus more interesting to ask the question if, starting from a fermionic Gaussian Hamiltonian, there are cases where the ground state cannot be adiabatically deformed to the atomic limit with a trivial entanglement bound and an integer particle number. The answer is, of course, yes. For an insulator in a symmetry-protected topological phase, such a deformation to a trivial state is not possible without closing the gap or breaking the symmetry. The tools we have developed so far, can thus be used to establish non-trivial lower bounds for the symmetry-resolved entanglement. We will discuss some examples in the next section. 

\section{Symmetry-protected topological phases}
\label{section:CN}
In symmetry-protected topological phases, the ground state cannot adiabatically be connected to the atomic limit without closing the excitation gap or breaking the symmetry. There are, broadly speaking, two types of symmetry-protected topological order: (i) Order protected by non-spatial symmetries, in particular time reversal, charge conjugation, and chiral symmetry, which leads to the tenfold classification \cite{tenfold1,tenfold2}. (ii) Spatial symmetries such as inversion, mirror, or rotational symmetries. In the latter case, one often speaks about topological crystalline insulators. In both cases we expect that in a topological non-trivial phase, there is a non-trivial minimal bound for the entanglement components. Here we want to provide some examples how the methods developed in the previous sections can be used to obtain bounds both for topological phases phases protected by non-spatial and by spatial symmetries.

\subsection{Chiral Symmetry}
A bilinear fermionic system with chiral symmetry can be written as
\begin{equation}
\label{chiral}
    H=\sum_k (\Psi_k^{a \dag}\; \Psi_k^{b \dag} )\l(\begin{array}{cc} 0 & h_k \\ h_k^\dag & 0  \end{array}\r)\l(\begin{array}{c} \Psi^a_k \\ \Psi^b_k \end{array}\r)
\end{equation}
where $\Psi^a_k=(a^1_k\dots a^N_k)$ and similarly for $\Psi_k^b$. I.e., each unit cell has $2N$ elements with $N$ elements belonging to sublattice $A$ and $N$ elements belonging to sublattice $B$. Due to the chiral symmetry, hopping occurs only between $a$ and $b$ elements, leading to the off-diagonal structure of the Hamiltonian matrix in Eq.~\eqref{chiral}. The topological phases of such a gapped system can be characterized by a winding number $\mathcal{I}\in\mathbb{Z}$.

For this system with periodic boundary conditions, we have recently proven that a topologically non-trivial phase, $\mathcal{I}\neq 0$, has $2|\mathcal{I}|$ protected eigenvalues at $1/2$ in the spectrum of the correlation matrix which is sometimes also called the single-particle entanglement spectrum \cite{MonkmanSirkerSpectrum}. From this, we concluded that the entanglement entropy has a non-trivial lower bound $S\geq 2|\mathcal{I}|\ln 2$. Since the Hamiltonian \eqref{chiral} does conserve the particle number, the entropy can be separated into number and configurational entropy. With the methods from the previous sections, we can now also obtain bounds for these symmetry-resolved components. Eq.~\eqref{RDM_blocks}, in particular, tells us that we can derive a bound based on the protected eigenvalues of the correlation matrix at $1/2$ alone. This is then essentially the case of Eq.~\eqref{Smax_general} with $M=2|\mathcal{I}|$. I.e., the bounds are
\begin{eqnarray}
    \label{chiral_bound}
    S&\geq&2|\mathcal{I}|\ln 2,\;\; S_c\geq\frac{1}{2^{2|\mathcal{I}|}}\sum_{n=0}^{2|\mathcal{I}|}\binom{2|\mathcal{I}|}{n}\ln\l[\binom{2|\mathcal{I}|}{n}\r] \nonumber \\ 
    S_N&\geq&2|\mathcal{I}|\ln 2-\frac{1}{2^{2|\mathcal{I}|}}\sum_{n=0}^{2|\mathcal{I}|}\binom{2|\mathcal{I}|}{n}\ln\l[\binom{2|\mathcal{I}|}{n}\r]
\end{eqnarray}
with the additional condition $S=S_N+S_c$, meaning that the three bounds are not all independent of each other.

\subsection{$C_n$-symmetric insulators}
Next, we want to discuss an example for a spatial, crystalline symmetry leading to topologically non-trivial phases. Let $\hat C_n$ be a generator of the cyclic group $\mathbb{Z}_n$. $\hat C_n$ is a conserved operator, acting on a particular Hilbert space, which is
unitary, Hermitian and which fulfills $(\hat C_n)^n=1$. 
Suppose that we have a non-interacting, gapped Hamiltonian $\hat H$ with a $\hat C_n$ symmetry i.e. $\hat{C_n} \hat{H} \hat{C_n}^{-1}=\hat{H}$. Each single particle eigenstate $|E_j\rangle$ of $\hat H$ is then also an eigenstate of the $\hat{C_n}$ operator. That is, for some integer $\ell_j$, $\hat{C_n} |E_j \rangle = e^{2\pi i \ell_j / n} |E_j \rangle$. The phase $e^{2\pi i \ell_j / n}$ is called the angular momentum of the state, labelled by quantum numbers $\ell_j$. For fixed particle  $N$, the normalized, many-particle ground state $|\Psi \rangle$ of the Hamiltonian has a $\mathbb{Z}^n$ invariant $Z=(Z_1,Z_2,\dots,Z_n)$, where $Z_j$ is the number of filled states with angular momentum quantum number $\ell_j$. An insulating ground state can then be described by 
\begin{equation}
\label{Psi}
    |\Psi\rangle = \prod_{j=1}^{Z_1}|Z_1,j \rangle \prod_{j=1}^{Z_2}|Z_2,j \rangle \dots \prod_{j=1}^{Z_n}|Z_n,j \rangle 
\end{equation}
where each $|Z_r , j \rangle$ state describes an orthogonal single particle state of momentum $\ell_r$ and energy $E_{r,j}$. The total number of particles is then $N=\sum_{i=1}^n Z_i$. 

\subsubsection{$C_2$ symmetry:}
The simplest case is that of a $C_2$ symmetry which, in a lattice system, could for example be due to an inversion or a mirror symmetry. In this case $C_2 |E_j \rangle = \pm |E_j \rangle$, i.e., single particle eigenstates are either even or odd and the numbers $Z_{1,2}$ of filled symmetric and antisymmetric states are topological invariants. The relevant quantity for the protected entanglement obtained from a reduced density matrix when the system is cut into to equal halves is $\Delta=|Z_1-Z_2|$. In Ref.~\cite{C2MonkmanSirker} we have recently proven---directly based on properties of the ground state and independent of the correlation matrix spectrum---that for $\Delta\neq 0$, non-trivial lower bounds for $S$, $S_N$, and $S_c$ exist. With the methods developed in this article, we can actually rederive this result in a different way. From Ref.~\cite{Bernevig2013} it is known that a $C_2$ symmetric system which is cut in half has $\Delta$ protected eigenvalues of the correlation matrix at $1/2$. Thus, we can use the same arguments as in the chiral case and the bounds from Ref.~\cite{C2MonkmanSirker} are reproduced by replacing $2|\mathcal{I}|\to\Delta$ in Eq.~\eqref{chiral_bound}.

\subsubsection{The general $C_n$ case:}
In the $C_n$ case there are, in general, multiple cuts of the system which are consistent with the symmetry. We will specify these cuts by first defining integers $m_1$ and $m_2$ such that both $m_1$ and $m_2$ divide $n$. The cut, denoted as $A^{m_2}_{1/m_1}$, will divide the system into subsystems $A$ and $B$. Let $M$ be the total number of states in the subsystem $A^{m_2}_{1/m_1}$. We require $A^{m_2}_{1/m_1}$ to have the following properties: (a) The number of single-particle states in $A$ is $1/m_{1}$ times the total number of states. (b) The subspace $A$ has a $\hat C_{m_2}$ symmetry. That is, there is a $\hat C_{m_2}$ symmetry that maps all single-particle states $|A_i \rangle$ back into the same subspace $A$. (c) The single particle states in the total system can be generated by acting with the operator $\hat C_{m_1 m_2}$ on the $|A_i \rangle$ states.
\begin{table*}[ht!] 
 \renewcommand{\arraystretch}{0.3}
\begin{tabularx}{1.0\columnwidth}{ p{0.9cm} p{0.9cm} p{0.9cm} p{1.0cm} p{7.5cm} p{3cm}}
 $n$ & $m_1$ & $m_2$ & \centering cut & \centering $\Delta$ & \ \ \ \ $p_n$ \\
 \hline \\
 2  & 2  & 1 &  \begin{minipage}{.1\textwidth} \includegraphics[width=0.5\linewidth]{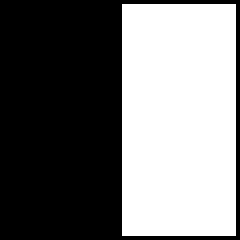} \end{minipage}  & \centering \small $|Z_1-Z_2|$ & \ \ \ \  \small $\dfrac{B_n^\Delta}{2^\Delta}$ \\ \\
  3  & 3  & 1 & \begin{minipage}{.1\textwidth} \includegraphics[width=0.5\linewidth]{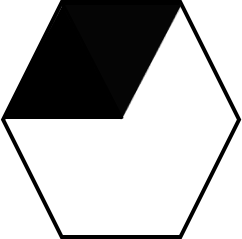} \end{minipage}  & \centering \begin{minipage}{.3\textwidth} \small $\max_{i,j=1,2,3} |Z_i-Z_j|$ \end{minipage} &  \small $\dfrac{2^{\Delta-n} \ B_n^\Delta}{3^\Delta}$ \\ \\
 4  & 2  & 1 & \begin{minipage}{.1\textwidth} \includegraphics[width=0.5\linewidth]{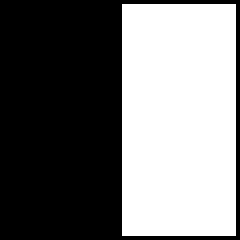} \end{minipage}  & \centering \small $|Z_1+Z_3-Z_2-Z_4|$ &  \ \ \ \ \small $\dfrac{B_n^\Delta}{2^\Delta}$ \\ \\
  4  & 2  & 2 & \begin{minipage}{.1\textwidth} \includegraphics[width=0.5\linewidth]{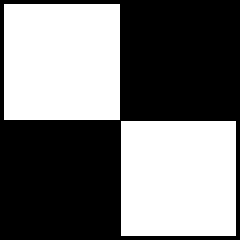} \end{minipage}  & \centering \small $|Z_1-Z_3|+|Z_2-Z_4|$ & \ \ \ \  \small $\dfrac{B_n^\Delta}{2^\Delta}$ \\ \\
  4  & 4  & 1 & \begin{minipage}{.1\textwidth} \includegraphics[width=0.5\linewidth]{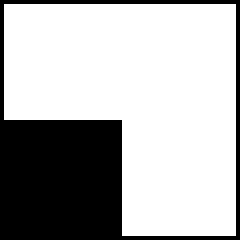} \end{minipage}  & \centering \begin{minipage}{.3\textwidth} \centering \small $\max_{i,j=1,2,3} |Z_i-Z_j|$ \end{minipage} &  \small $\dfrac{3^{\Delta-n} \ B_n^\Delta}{4^\Delta}$ \\ \\
  6  & 2  & 1 & \begin{minipage}{.1\textwidth} \includegraphics[width=0.5\linewidth]{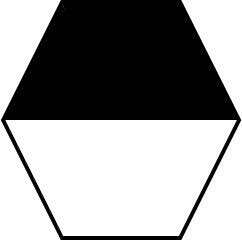} \end{minipage}  & \centering \begin{minipage}{.3\textwidth} \small $|Z_1+Z_3+Z_5-Z_2-Z_4-Z_6|$ \end{minipage} &  \ \ \ \  \small $\dfrac{B_n^\Delta}{2^\Delta}$ \\ \\
  6  & 2  & 3 & \begin{minipage}{.1\textwidth} \includegraphics[width=0.5\linewidth]{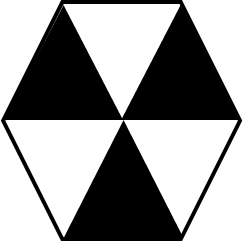} \end{minipage}  & \centering \begin{minipage}{.35\textwidth} \small $|Z_1-Z_4|+|Z_2-Z_5|+|Z_3-Z_6|$ \end{minipage} &  \ \ \ \  \small $\dfrac{B_n^\Delta}{2^\Delta}$ \\ \\
  6  & 3  & 1 & \begin{minipage}{.1\textwidth} \includegraphics[width=0.5\linewidth]{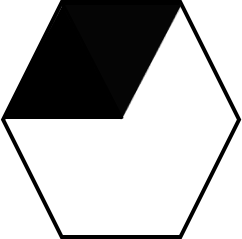} \end{minipage}  & \centering \begin{minipage}{.4\textwidth} \footnotesize $\max(|Z_1+Z_4-Z_2-Z_5|$,\\$|Z_1+Z_4-Z_3-Z_6|, |Z_2+Z_5-Z_3-Z_6|)$ \end{minipage} &  \small $\dfrac{2^{\Delta-n} \ B_n^\Delta}{3^\Delta}$ \\ \\
  6  & 3  & 2 & \begin{minipage}{.1\textwidth} \includegraphics[width=0.5\linewidth]{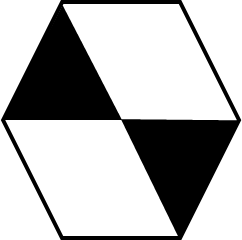} \end{minipage}  & \centering \begin{minipage}{.4\textwidth} \footnotesize $\max(|Z_1-Z_3|,|Z_1-Z_5|,|Z_3-Z_5|)$\\ $+\max(|Z_2-Z_4|,|Z_2-Z_6|,|Z_4-Z_6|)$ \end{minipage} &  \small $\dfrac{2^{\Delta-n} \ B_n^\Delta}{3^\Delta}$ \\ \\
  6  & 6  & 1 & \begin{minipage}{.1\textwidth} \includegraphics[width=0.5\linewidth]{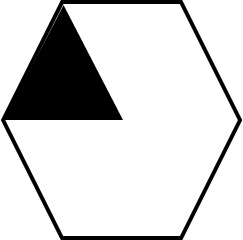} \end{minipage}  & \centering \begin{minipage}{.3\textwidth} \small $\max_{i,j=1,2,3,4,5,6} |Z_i-Z_j|$ \end{minipage} &  \small $\dfrac{5^{\Delta-n} \ B_n^\Delta}{6^\Delta}$ \\ \\
\end{tabularx}
\caption{Table taken from Ref.~\cite{Bernevig2013} and extended by the probabilities $p_n$ which give a lower bound on the number entropy $S_N$. $p_n$ and $B_n^\Delta$ are defined in Eq.~\eqref{pn}.}
\label{Table1}
\end{table*} 
We will now discuss two different methods to obtain bounds for the entanglement components.

\paragraph{Method 1:}
Here, we follow Ref.~\cite{Bernevig2013} where it was shown that for any symmetry-respecting cut of a $C_n$ symmetric system, there are protected eigenvalues in the spectrum of the correlation matrix. In particular, it was found that there are $\Delta$ eigenvalues in the range $[1/{m_1},1-1/{m_1}]$, where the $\Delta$ values for various symmetries and cuts are given in table \ref{Table1}. Using the results derived here, we can obtain bounds for the entanglement components based on these protected eigenvalues alone. In the case $m_1=2$ (system is cut in half), we have $\Delta$ protected eigenvalues exactly at $1/2$. In this case we obtain again the bounds \eqref{chiral_bound} with $2|\mathcal{I}|\to\Delta$ and $\Delta$ given in table \ref{Table1}.

\begin{figure}
\centering
\includegraphics[width=0.8\textwidth]{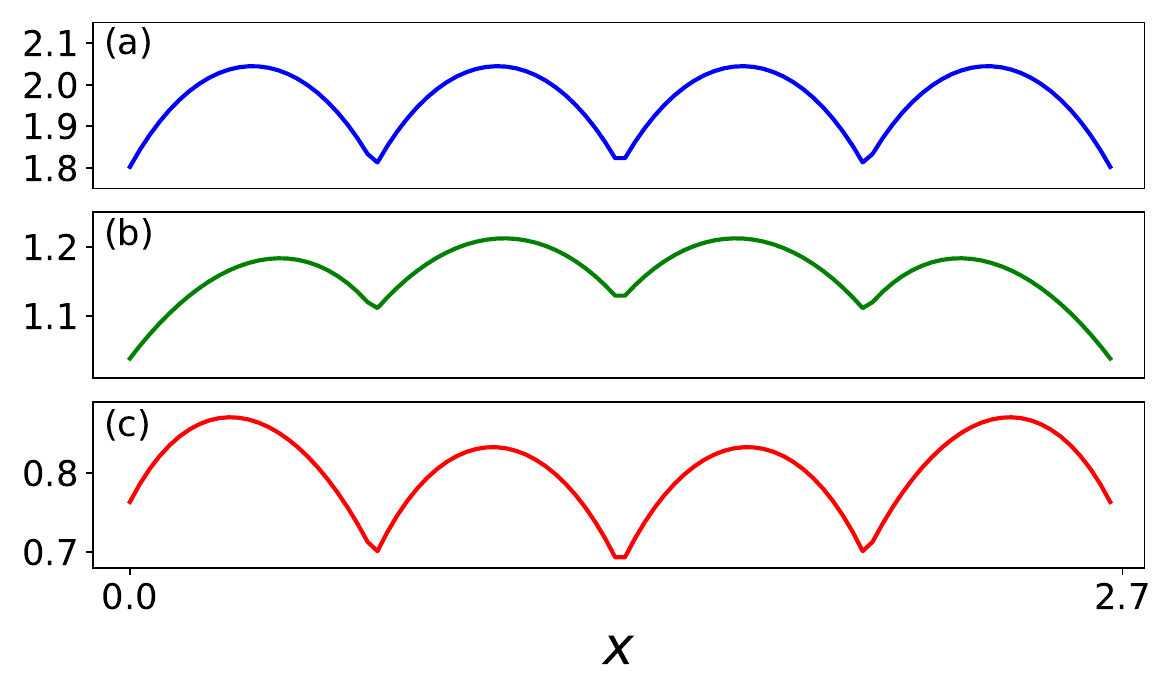} 
\caption{Minimal entropies for $\Delta=4$, $m_1=6$ as a function of $x$ using method 1. (a) Entropy $S$, (b) Number entropy $S_N$, and (c) Configurational entropy $S_c$. }
\label{fig:method1}
\end{figure}

When $m_1 \neq2$, we cannot simply apply majorization to the protected eigenvalues in the range $[1/{m_1},1-1/{m_1}]$ since the sum of these eigenvalues is not fixed. What we can do, however, is vary the sum of the eigenvalues and consider the maximally majorized entropy for each fixed sum. For real numbers $x$, $a$ and $b$, we define the function
\begin{equation}
    F[x,a,b] = \begin{cases}
        0 & x<a \\
        x-a & a \leq x \leq b \\
        b-a & b<x.
    \end{cases}
\end{equation}
 We can parameterize the eigenvalue sum as $\sum_{i=1}^\Delta C_i=\frac{\Delta}{m_1}+x$. Let $\mu=(1-\frac{2}{m_1})$. Then, for a fixed $x$, the maximal majorization occurs when 
\begin{equation}
\label{majorization}
    C_i = 1/{m_1}+F[x,(i-1)\mu,i\mu]\, .
\end{equation}
The average particle number in the subsystem is given by $\langle N\rangle = \sum_{i=1}^\Delta C_i + \sum_{\Delta+1}^M C_i$, i.e., it is given by a sum of the protected and unprotected eigenvalues. If we assume that we can always obtain the given average particle number while freely varying the sum of the protected eigenvalues, we can then ask the question what the global minimum as a function of $x$ is if we choose the $C_i$ giving maximal majorization for each $x$ according to Eq.~\eqref{majorization}. An example for $\Delta=4$ and $m_1=6$ is shown in Fig.~\ref{fig:method1}. 

The local minima in all three entanglement entropies correspond to having all the $C_i$ at the boundaries of the interval $[1/{m_1},1-1/{m_1}]$. This can be understood by recalling that all three entropy components are concave functions in the $\{ C_i\}$. The question then becomes for which boundary configuration the entropies become minimal. We note that if $C_i=\frac{1}{m_1}$ then $(1-C_i)=1-\frac{1}{m_1}$. According to Eq.~\eqref{SvN}, we therefore get exactly the same contribution to the entanglement entropy $S(\{C_i\})$ whether an eigenvalue sits at the lower or upper boundary. Thus, any arrangement of the $\{C_i\}$ at the two boundaries of the interval leads to the same minimum. 

For the number entropy $S_N$, the two cases where all the eigenvalues sit at $\frac{1}{m_1}$ or all the eigenvalues sit at $1-\frac{1}{m_1}$ will yield the same result and correspond, physically, to counting either particles or holes. If we put the eigenvalues $C_1,\dots,C_{\Delta-1}$ at $\frac{1}{m_1}$ then $S_N$ is a function of $C_\Delta$ with a minimum at $\frac{1}{m_1}$. We thus conclude that the global minimum of $S_N$ is obtained if all the $C_i$ sit at $\frac{1}{m_1}$ or all the $C_i$ sit at $1-\frac{1}{m_1}$. For this case, we can obtain a closed form expression for the probabilities $p_n$ to have $n$ particles in the subsystem. In order to have $n$ particles we need $n$ factors of $C_i=\frac{1}{m_1}$ and $\Delta-n$ factors of $1-C_i=\frac{m_1-1}{m_1}$. There are $\binom{\Delta}{n}$ possible combinations. Thus
\begin{equation}
\label{pn}
p_n = \l(\frac{1}{m_1}\r)^n\l(\frac{m_1-1}{m_1}\r)^{\Delta-n}\binom{\Delta}{n} 
= \frac{(m_1-1)^{\Delta-n}}{m_1^\Delta} B_n^\Delta
\end{equation}
where we have again used $B_n^\Delta=\binom{\Delta}{n}$. This allows us to obtain a lower bound for the number entropy $S_N$ for any $C_n$ symmetry and any of the cuts shown in table \ref{Table1}. 

Finally, we consider also the minimum for the configurational entropy $S_c$. As can be seen in Fig.~\ref{fig:method1}, the global minimum does not occur for the same configuration of the $C_i$'s as for the number entropy. Based on the $M=2,3$ (equivalent to $\Delta=2,3$ in this context) cases analyzed earlier and several additional examples we considered numerically, we hypothesize that the global minimum for $S_c$ is obtained if $\Delta/2$ [$(\Delta\pm 1)/2$] eigenvalues are placed at $\frac{1}{m_1}$ and $\Delta/2$ [$(\Delta\mp 1)/2$] eigenvalues at $1-\frac{1}{m_1}$ for $\Delta$ even [odd]. While this method does provide lower bounds for the symmetry-resolved entanglement, these bounds are, in general, not optimal. We will demonstrate this next for one particular type of cut. 

\paragraph{Method 2:}
\begin{figure*}
\centering
\includegraphics[width=0.32\textwidth]{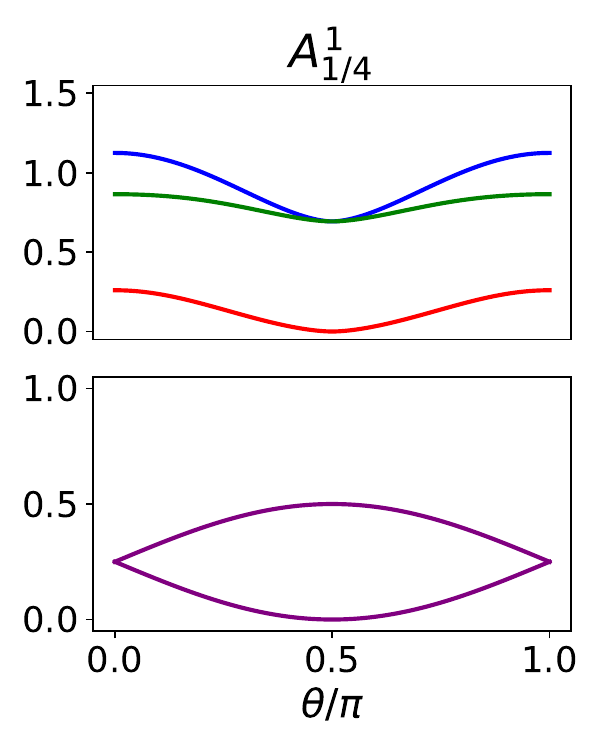} \ 
\includegraphics[width=0.32\textwidth]{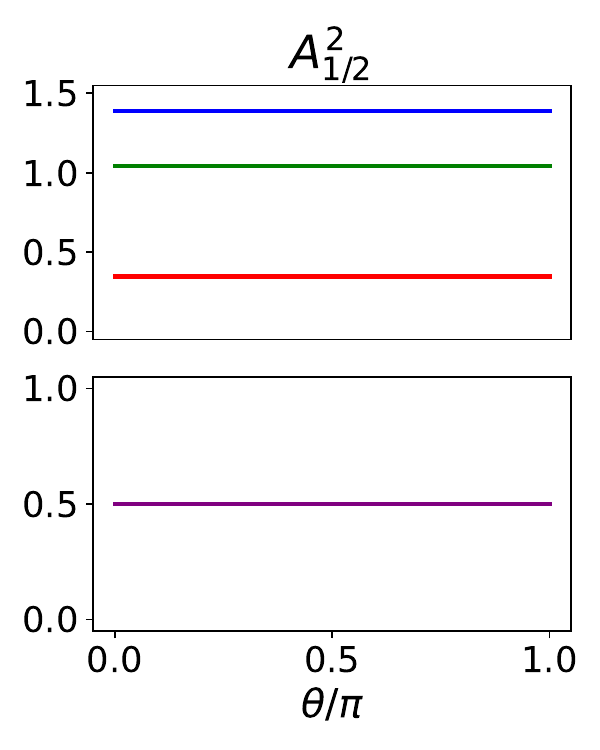} \ 
\includegraphics[width=0.32\textwidth]{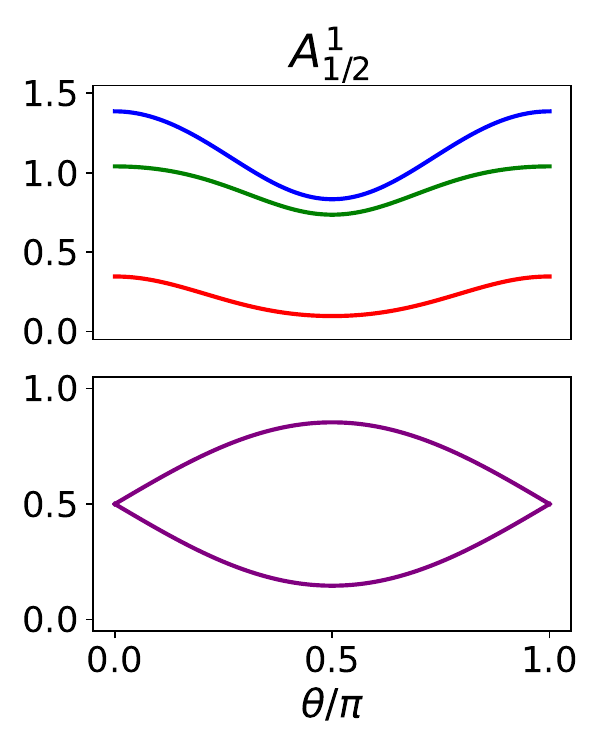}
\caption{Entanglement in the adiabatic deformation of the state \eqref{ex1}. Top row: Symmetry-resolved entanglement components for each cut. $S$ (blue), $S_N$ (green), $S_C$ (red). Bottom row: Single-particle entanglement spectrum for each cut.}
\label{fig:adiabaticFigure}
\end{figure*}
Consider the cut $A_{1/n}^1$, i.e., $m_1=n$ and the subsystem is the $1/n$-th part of the system. Then the state is given by Eq.~\eqref{Psi} and the reduced single-particle correlation matrix can be written as
\begin{equation}
    C = C(Z_1)+C(Z_2)+\dots+C(Z_n)
\end{equation}
where $C(Z_r)$ is the reduced correlation matrix of the state $\prod_{j=1}^{Z_r}|Z_r,j \rangle$. These states are orthogonal and there are $Z_r$ eigenvalues at $1/n$.

Now we define a vector $\tilde{\vec{Z}}=(\tilde{Z}_{1},\tilde{Z}_{2},\dots,\tilde{Z}_{n})$ which is an ordered version of the invariant $\vec{Z}$. That is, the values $\tilde{Z}_{j}$ are a reordering of $Z_j$ such that $\tilde{Z}_{1} \geq \tilde{Z}_{2} \geq \dots \geq \tilde{Z}_{n}$. Then we define invariant differences $\Delta_{1}=(\tilde{Z}_{1}-\tilde{Z}_{2})$, $\Delta_{2}=(\tilde{Z}_{2}-\tilde{Z}_{3})$, $\dots$, $\Delta_{n-1}=(\tilde{Z}_{n-1}-\tilde{Z}_{n})$, $\Delta_{n}=\tilde{Z}_{n}$. Next, we define a non-increasing vector $\vec{x}$ with $\Delta_r$ values at $r/n$ for $r=1,2,\dots,n$ and the remaining values equal to zero. Then, applying the majorization theorem \cite{fulton1998eigenvalues} for sums of Hermitian matrices, we find that all sets of possible eigenvalues of $C$ are majorized by $\vec{x}$. 

Thus, we find that the state with minimal $S$, $S_N$, and $S_c$ has $\Delta_r$ eigenvalues at $r/n$ for $r=1,2,\dots,n$. This comes to a total of $(\tilde{Z}_1 - \tilde{Z}_n)$ eigenvalues which are non-zero. The previous known minimum for $S$ \cite{Bernevig2013} is also known to have $(\tilde{Z}_1 - \tilde{Z}_n)$ non-zero eigenvalues but at a value of $\frac{1}{n}$ or $\frac{n-1}{n}$. Thus, by using majorization, we have found in this particular case stronger lower entropy bounds than previously known. In fact, because we consider all the eigenvalues of the correlation matrix and $\vec{x}$ majorizes all possible arrangements of these eigenvalues we know that this bound is optimal. For cuts besides the $A_{1/n}^{1}$ cut, this majorization method is not directly applicable. 
\begin{figure*}
\centering
\begin{tabular}{c@{\hspace{0.1\textwidth}}c@{\hspace{0.1\textwidth}}}
     \includegraphics[width=0.4\textwidth,valign=m]{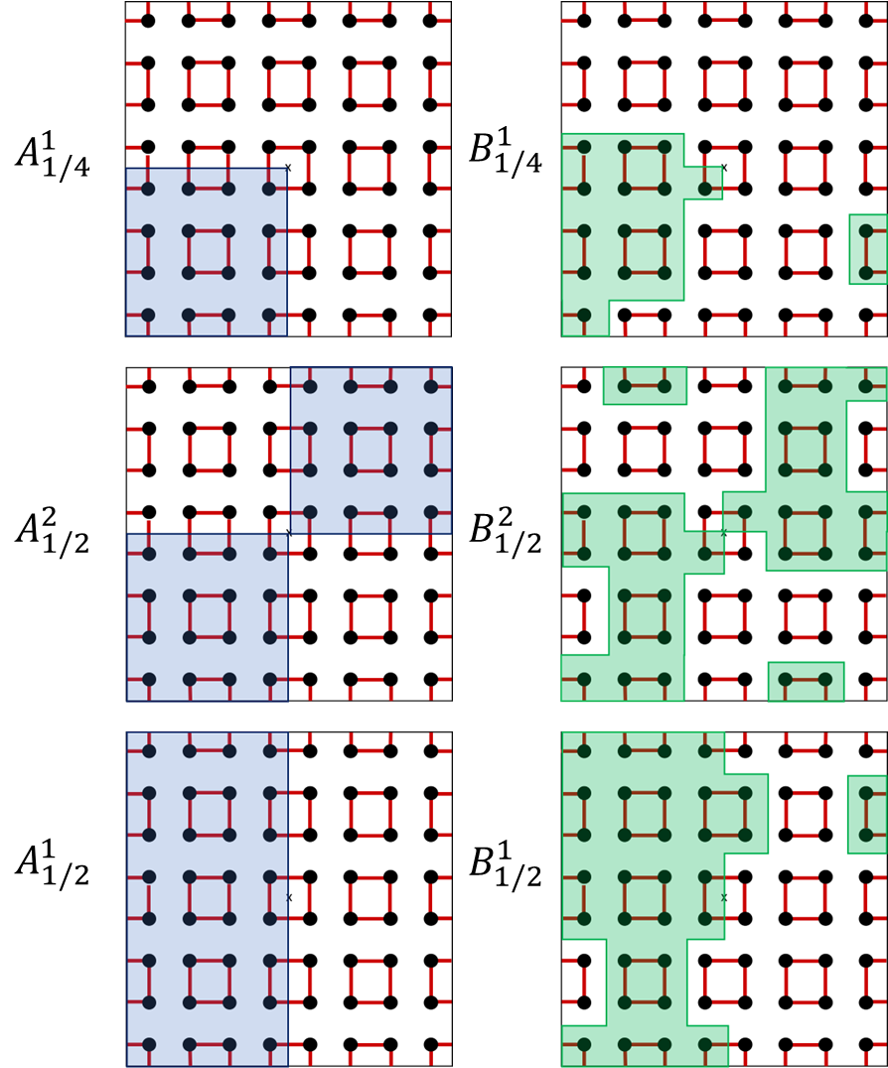} &  \includegraphics[width=0.5\textwidth,valign=m]{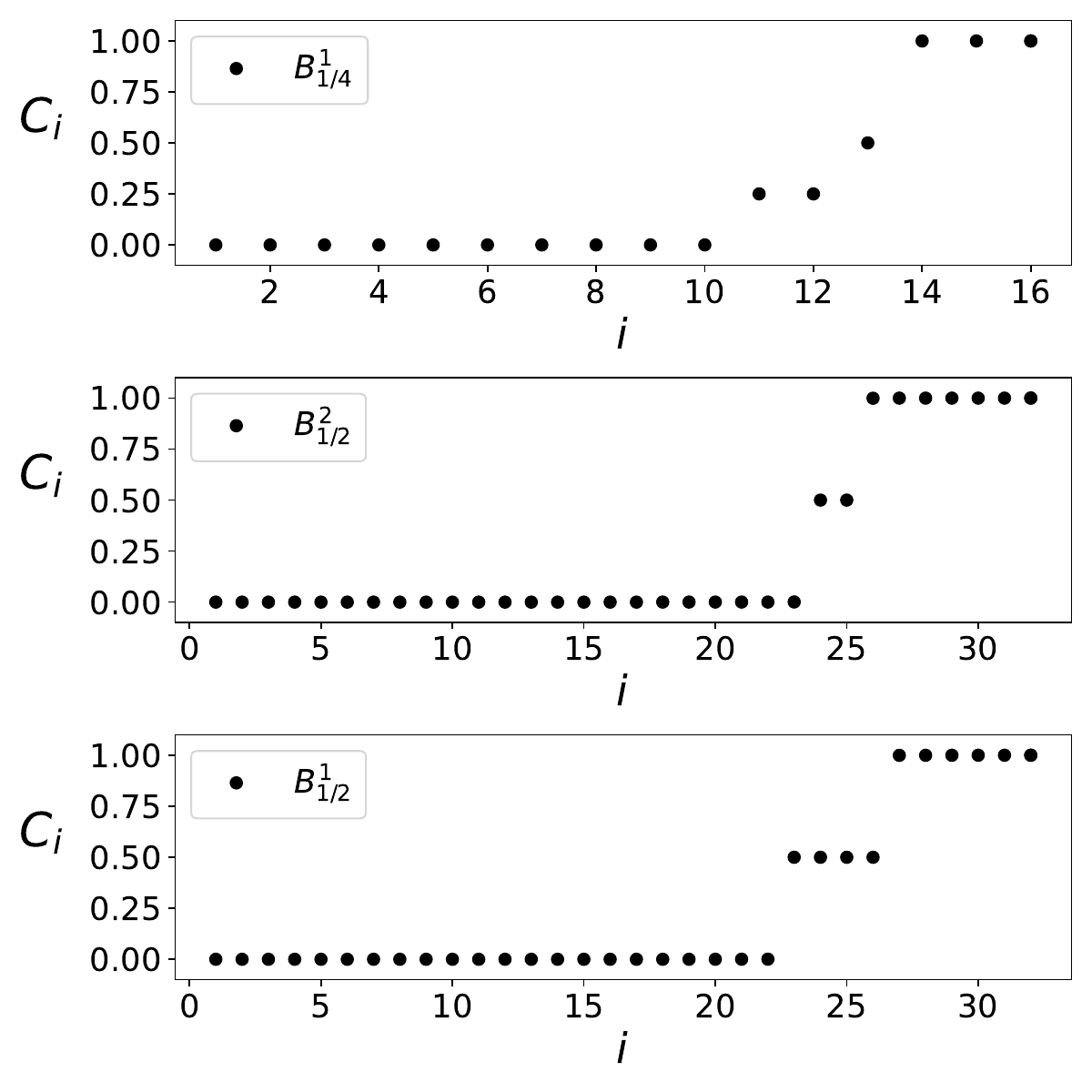} \\
\end{tabular}
\caption{Left: We consider two types of cuts, $A$ and $B$, for each cut classification. The $B$ cuts result in minimal entanglement for this $C_4$-symmetric system. Right: Single-particle entanglement spectra with the eigenvalues ordered by magnitude for the $B$ cuts with hopping amplitudes $t_1=0.1$ and $t_2=1.9$.}
\label{example2}
\end{figure*}

\section{$C_4$ symmetry examples}
\label{section:CNexamples}
In this section, we consider two examples for fermionic Gaussian systems with $C_4$ symmetry. In the first example, we will demonstrate an adiabatic deformation of the Hamiltonian and monitor all aspects of entanglement including the symmetry-resolved components and the single-particle entanglement spectrum. In the second example, we will consider a two-dimensional plaquette model and show that by choosing specific cuts, consistent with the $C_4$ symmetry, the entanglement entropy can be reduced to the lower bounds found earlier.

\subsection{Example 1: Adiabatic Deformation}
First, we will consider an $C_4$ symmetric example with invariant $\vec{Z}=(1,1,0,0)$. Applying method 1 from the previous section for a $A_{1/4}^{1}$ cut, we have $\Delta=1$, see table \ref{Table1}. Thus, there is one protected eigenvalue in the range $[\frac{1}{4},\frac{3}{4}]$. The lower bounds are then obtained by placing this eigenvalue at $\frac{1}{4}$. We will see from the solution of the example that this bound is not optimal. For the $A_{1/2}^{2}$ cut, we have $\Delta=2$ and two eigenvalues protected at $\frac{1}{2}$. This bound is optimal. For the $A_{1/2}^{1}$ cut, we have $\Delta=0$ and method 1 places no restrictions on the eigenvalues, i.e., the lower bounds are trivial. Using method 2, we can only consider the $A_{1/4}^1$ cut. This method predicts lower bounds for all entanglement components which are obtained from a single eigenvalue at $C_1=\frac{1}{2}$, which results in optimal bounds. 

Let us now describe the first example. Consider a system of two stacked $2\times 2$ plaquettes, i.e., a system consisting of a total of 8 sites. Operators acting on the lower plaquette are denoted by $a_L$, $b_L$, $c_L$, $d_L$ and those acting on the upper plaquette by $a_U$, $b_U$, $c_U$, $d_U$. We now define the following operators
\begin{eqnarray}
    \ell_j^{1} =\frac{a_j+b_j+c_j+d_j}{2} \ &,& \ \ell_j^{2} =\frac{a_j+\im b_j-c_j-\im d_j}{2} \nonumber \\
    \ell_j^{3} =\frac{a_j-b_j+c_j-d_j}{2} \ &,& \ \ell_j^{4} =\frac{a_j-\im b_j-c_j+\im d_j}{2}
\end{eqnarray}
for $j=L,U$. The $\hat{C}_4$ symmetry operator maps $a_j \rightarrow b_j$, $b_j \rightarrow c_j$, $c_j \rightarrow d_j$, $d_j \rightarrow a_j$. The Hamiltonian we want to consider is given by
\begin{equation}
    \hat{H}(\theta)=(-{\ell_{L}^1}^\dag \ell_{L}^1 + {\ell_U^{1}}^\dag \ell_U^1)  
    +(\cos 2\theta)({\ell_L^{2}}^\dag \ell_L^{2} - {\ell_U^{2}}^\dag \ell_U^{2}) + (\sin 2\theta) ({\ell_L^{2}}^\dag \ell_U^{2} + {\ell_U^{2}}^\dag \ell_L^{2}).
\end{equation}
The 2-particle ground state is given by
\begin{equation}
\label{ex1}
    |\Psi \rangle = {\ell_L^1}^\dag \ ( {\ell_L^2}^\dag \sin \theta- {\ell_U^2}^\dag \cos \theta) |0\rangle 
\end{equation}
where $|0\rangle$ is the vaccuum state and does depend on the parameter $\theta$. 

The $A_{1/4}^1$ cut consists of both $a_j$ lattice points. The $A_{1/2}^2$ cut consists of both $a_j$ and $c_j$ lattice points. Lastly, the $A_{1/2}^1$ cut consists of both $a_j$ and $b_j$ lattice points. The single-particle entanglement spectrum for each cut is given by
\begin{eqnarray}
    A_{1/4}^1&:& \ C_1 = \frac{1-\sin \theta}{4} \ , \ C_2 = \frac{1+\sin \theta}{4} \nonumber \\
    A_{1/2}^2&:& \ C_1 = \frac{1}{2} \ , \ C_2 = \frac{1}{2} \\
    A_{1/2}^1&:& \ C_1 = \frac{2-\sqrt{2} \sin \theta}{4} \ , \ C_2 = \frac{2+\sqrt{2}\sin \theta}{4}. \nonumber 
\end{eqnarray}
We can now use the formulas \eqref{M2_case} to obtain the eigenvalues of the reduced density matrix and the symmetry-resolved entanglement. Using these formulas, we find that all three symmetry-resolved entanglement components are minimized at the spectrum limits. This occurs when $\theta=\pi/2$ as shown in Fig.~\ref{fig:adiabaticFigure}. For the $A^1_{1/4}$ cut we see that method 2 indeed gives optimal lower bounds of $S=S_N=\ln 2$ and $S_c=0$. For the $A^2_{1/2}$ cut the two eigenvalues are always fixed at $1/2$ and method 1 gives optimal lower bounds which correspond to those values given in Eq.~\eqref{Smax_M2}. In the $A^1_{1/2}$ case method 1 gives only trivial bounds and method 2 is not applicable. We find, however, a very similar picture to the $A^1_{1/4}$ cut with the two eigenvalues coupled to each other and non-trivial lower entanglement values obtained for $\theta=\pi/2$.  

\subsection{Example 2: Selection of the cut}
As our second example, we will consider again a $C_4$-symmetric system but this time with invariant $\vec{Z}=(6,3,4,3)$. Applying method 1 for the $A_{1/4}^{1}$ cut, we have $\Delta=3$, see table \ref{Table1}. This predicts that three eigenvalues are protected in the range $[\frac{1}{4},\frac{3}{4}]$. The lower bounds for $S$ and $S_N$ are thus obtained for $C=(1/4,1/4,1/4)$ while the lower bound for $S_c$ is obtained for $C=(1/4,1/4,3/4)$. We will see that these bounds are not optimal. For the $A_{1/2}^{2}$ cut, we have $\Delta=2$ and two eigenvalues protected at $\frac{1}{2}$. For the $A_{1/2}^{1}$ cut, we have $\Delta=4$ and four eigenvalues protected at $\frac{1}{2}$. In the latter two cases the bounds will be optimal. Using method 2, we can improve on the bounds for the $A_{1/4}^1$ cut. This method predicts that better lower bounds for all entanglement components are obtained for $C=(1/4,1/4,1/2)$.

In this example, we will show that the lower bounds can be reached not only by an adiabatic deformation of the Hamiltonian as in the first example, but also by a deformation of the cut. The $A_{1/m_1}^{m_2}$ cuts shown in Fig.~\ref{example2} are the most natural cuts but lead to entanglement which is larger than the bounds. If we, however, deform these cuts to $B_{1/m_1}^{m_2}$, also shown in Fig.~\ref{example2}, then the single particle entanglement spectra correspond to those used to obtain the lower bounds.

The fermionic model we will consider consists of a unit cell with operators $a,b,c,d$ and is given by 
\begin{eqnarray}
\hat{H} &=& t_1\sum_{x,y} (a_{x,y}^\dag b_{x,y} + a_{x,y}^\dag c_{x,y}+b_{x,y}^\dag d_{x,y}+c_{x,y}^\dag d_{x,y}) \nonumber \\ 
&+& t_2 \sum_{x,y} \l(b_{x,y}^\dag a_{x+1,y} + c_{x,y}^\dag a_{x,y+1}\r. 
+ \l.d_{x,y}^\dag b_{x,y+1}+d_{x,y}^\dag c_{x+1,y}\r) + \textit{h.c.} 
\end{eqnarray}
where $x,y$ enumerate the unit cells. Here we will consider the case of a two-dimensional system with $4\times 4$ unit cells (plaquettes) with $16$ fermions where $t_1=0.1$ and $t_2=1.9$. Shown in Fig.~\ref{example2} are only the strong bonds $t_2$ for various cuts. For the two cases $A^1_{1/2}$ and $A^2_{1/2}$, method 1 predicts optimal lower bounds which correspond to $\Delta$ eigenvalues fixed at $1/2$. For the case $B_{1/4}^1$ state, both method 1 and method 2 provide correct lower bounds for the entanglement entropy components but only method 2 provides the optimal bound corresponding to eigenvalues $C=\{1/4,1/4,1/2\}$, see Fig.~\ref{example2}.

\section{Conclusion}
In a system with particle number conservation, the entanglement entropy can be split into two components: the number entropy and the configuational entropy. The first main result we did establish in this paper is that while the entanglement entropy is extensive, the symmetry-resolved components are not and rather fulfill the inequalities \eqref{SCSNresult}. These inequalities are general and apply to any system, fermionic or bosonic, non-interacting or interacting, so long as particle number conservation is respected.

In the rest of the paper, we concentrated on obtaining rigorous results for the symmetry-resolved entanglement components in fermionic Gaussian systems. We introduced two techniques: First, we noticed that each eigenvalue of the correlation matrix can be treated as its own 'independent subsystem', thus allowing to obtain bounds based on the knowledge of a subset of these eigenvalues together with the inequalities \eqref{SCSNresult}. Second, we introduced majorization as a technique to obtain strict bounds on the entanglement components. To do so, one needs to prove first that the entanglement functions are concave in the eigenvalues of the correlation matrix $\{C_k\}$. This is obvious and well-known for the entanglement entropy. Here, we have proven using the Shepp-Olkin theorem that the number entropy $S_N$ is also concave in the $\{C_k\}$. For the configurational entropy we were not able to establish a full proof that this entropy is concave in the $\{C_k\}$ as well but hypothesize that it is based on the results for a small subsystems. Without further restrictions on the eigenvalues, majorization then allows to obtain upper bounds for all entanglement components for a given average filling of the subsystem.

It is physically even more interesting to apply these techniques to derive non-trivial lower bounds for systems which cannot be adiabatically connected to the atomic limit. This is, for example, the case for Gaussian fermionic systems with symmetry-protected insulating topological phases. As examples, we derived lower bounds for systems with a chiral symmetry as well as for systems with a spatial $C_n$ symmetry. In the latter case we discussed two methods. The first was based on results obtained in Ref.~\cite{Bernevig2013}. These results showed that in $C_n$ symmetric systems, a certain number of correlation matrix eigenvalues are restricted to a range $[1/m_1,1-1/m_1]$ where $1/m_1$ is the fraction of the total system the subsystem consists of. We could show that the lower bound for the number entropy is then obtained by placing all the eigenvalues at the same boundary while the lower bound for the configurational entropy is obtained by equally distributing the eigenvalues between the two boundary values. While these bounds are always valid, they are not necessarily optimal. We could show, in particular, in our second approach how for a specific cut better and, indeed, optimal bounds can be obtained. We illustrated these results by considering two concrete examples for $C_4$-symmetric systems.  

In the first example we considered, we have also seen that lower bounds for the entanglement components can exist even in cases where the methods used here only give trivial bounds. This clearly indicates that while some progress has been achieved here, the problem of giving optimal bounds for symmetry-protected topological phases is far from being fully solved. Another interesting problem is, of course, topological phases in interacting fermionic or bosonic systems. However, then the single-particle correlation matrix is no longer directly related to the reduced density matrix and completely new and different tools need to be developed to establish bounds for the symmetry-resolved entanglement.

\section*{Acknowledgment}
The authors acknowledge support by the Natural Sciences and Engineering Research Council (NSERC, Canada). K.M. acknowledges support by the Vanier Canada Graduate Scholarships Program. J.S. acknowledges by the Deutsche Forschungsgemeinschaft (DFG) via Research Unit FOR 2316. K.M. would like to thank A. Urichuk for helpful discussions.

\section*{References}
\providecommand{\newblock}{}

\end{document}